\documentclass[conference]{IEEEtran}
\IEEEoverridecommandlockouts
\usepackage{cite}
\usepackage{amsmath,amssymb,amsfonts}
\usepackage{algorithmic}
\usepackage{graphicx}
\usepackage{textcomp}
\usepackage{xcolor}
\usepackage{balance}
\usepackage{soul}
\def\BibTeX{{\rm B\kern-.05em{\sc i\kern-.025em b}\kern-.08em
    T\kern-.1667em\lower.7ex\hbox{E}\kern-.125emX}}

\usepackage{caption, subcaption}

\begin{document}

\title{Circuit Partitioning for Multi-Core Quantum Architectures with Deep Reinforcement Learning
 \thanks{Authors acknowledge support from the EU's Horizon Europe program through the European Research Council (ERC) under grant agreement 101042080 (WINC) and through the European Innovation Council (EIC) PATHFINDER scheme, grant agreement No 101099697 (QUADRATURE).}
 }

% Author template
%  \author{\IEEEauthorblockN{1\textsuperscript{st} Given Name Surname}
% \IEEEauthorblockA{\textit{dept. name of organization (of Aff.)} \\
% \textit{name of organization (of Aff.)}\\
% City, Country \\
% email address or ORCID}
% \and

% Centrar autors segona fila: https://tex.stackexchange.com/questions/458204/ieeetran-document-class-how-to-align-five-authors-properly/458208#458208
% \makeatletter
% \newcommand{\linebreakand}{%
%   \end{@IEEEauthorhalign}
%   \hfill\mbox{}\par
%   \mbox{}\hfill\begin{@IEEEauthorhalign}
% }
% \makeatother

\author{\IEEEauthorblockN{Arnau Pastor}
\IEEEauthorblockA{\textit{Universitat Politècnica de Catalunya}\\
Barcelona, Spain \\
arnau.pastor.lacueva@estudiantat.upc.edu}
\and
\IEEEauthorblockN{Pau Escofet}
\IEEEauthorblockA{\textit{Universitat Politècnica de Catalunya}\\
Barcelona, Spain \\
pau.escofet@upc.edu}
\and
\IEEEauthorblockN{Sahar Ben Rached}
\IEEEauthorblockA{\textit{Universitat Politècnica de Catalunya}\\
Barcelona, Spain \\
sahar.benrached@upc.edu}
% \linebreakand % <------------- \and with a line-break
\and
\IEEEauthorblockN{Eduard Alarcón}
\IEEEauthorblockA{\textit{Universitat Politècnica de Catalunya}\\
Barcelona, Spain \\
eduard.alarcon@upc.edu}
\and
\IEEEauthorblockN{Pere Barlet-Ros}
\IEEEauthorblockA{\textit{Universitat Politècnica de Catalunya}\\
Barcelona, Spain \\
pere.barlet@upc.edu}
\and
\IEEEauthorblockN{Sergi Abadal}
\IEEEauthorblockA{\textit{Universitat Politècnica de Catalunya}\\
Barcelona, Spain \\
abadal@ac.upc.edu}
}

\maketitle
\begin{abstract}
Quantum computing holds immense potential for solving classically intractable problems by leveraging the unique properties of quantum mechanics. The scalability of quantum architectures remains a significant challenge. Multi-core quantum architectures are proposed to solve the scalability problem, arising a new set of challenges in hardware, communications and compilation, among others. One of these challenges is to adapt a quantum algorithm to fit within the different cores of the quantum computer. This paper presents a novel approach for circuit partitioning using Deep Reinforcement Learning, contributing to the advancement of both quantum computing and graph partitioning. This work is the first step in integrating Deep Reinforcement Learning techniques into Quantum Circuit Mapping, opening the door to a new paradigm of solutions to such problems.
% By addressing the challenges associated with scaling quantum computers, we pave the way for their practical implementation in solving computationally challenging problems.
\end{abstract}


\section{Introduction}
Quantum computing is a rapidly evolving field that has the potential to revolutionize computation, making certain types of classically intractable problems solvable by leveraging quantum mechanics properties such as superposition or entanglement. Its applications range from prime factorization and simulations to healthcare and finances  \cite{shor, grover, low, peruzzo, woerner, rasool}.
% It has a wide range of applications, including prime factorization, physics and chemistry simulation, database search and finances, among others \cite{shor, grover, low, peruzzo, woerner, rasool}.

Current quantum computers use diverse qubit technologies \cite{superconducting, photonic, dots, ions}, not exceeding a few thousand qubits \cite{gambetta}, a much lower number than the qubits required for practical quantum applications \cite{nisq}. Increasing the number of qubits within a single quantum processor comes with huge technical challenges, such as cross-talk, disturbances in quantum states, and increased complexity in control systems. Therefore, scaling quantum computers within these monolithic architectures remains a formidable task.

An alternative to single-core quantum computers are modular (or multi-core) quantum computers, hosting several quantum cores connected among them to increase the number of qubits within the whole system \cite{jnane, santiago, smith}. The transfer of qubits between cores is accomplished through methods such as teleportation \cite{remote}, using EPR pairs \cite{epr} as communication resources, or quantum coherent inter-core communication \cite{gold}. 
%There exist several proposals on how to connect cores, such as interconnecting the cores with coupling-enabled links \cite{gold}, or interconnecting the cores by a photonic network that distributes quantum entangled states \cite{epr} to different cores, such states are used as communication resources, enabling quantum teleportation or remote gate execution \cite{santiago}. This work focuses on this last multi-core architecture model.
However, multi-core quantum computing architectures come with several challenges. 
This work focuses on executing a quantum circuit in a modular architecture, a problem known as quantum circuit mapping, which remains highly unexplored for multi-core architectures \cite{chong, qubo, hqa}. We will use the average number of qubit movements across cores as the evaluation metric, aiming at minimizing it.

This work proposes to use Deep Reinforcement Learning (DRL) to solve the mapping problem for multi-core architectures. The contributions of this paper can be summarized as follows:
\begin{itemize}
    \item The proposal of several DRL models to approach the mapping problem.
    \item A comparison between the proposed DRL approaches and a state-of-the-art solution.
\end{itemize}

The obtained results show that DRL stands as a viable approach for quantum circuit mapping. These findings not only underscore the potential of DRL but also encourage us to intensify the research efforts in this direction, inviting further exploration and innovation, ultimately advancing the frontiers of quantum technology and its applications.

The remainder of this paper is organized as follows. In Section \ref{sec:background}, a brief introduction to quantum circuit mapping and DRL is provided, allowing the reader to be familiar with the technologies used in this work. Section \ref{sec:DRL_Exploration} introduces several DRL models to solve the mapping problem and compares them. Using the proposed models, Section \ref{sec:Performance_Assessment} performs an assessment of the models, comparing them with the SOTA algorithms. Lastly, Section \ref{sec:Conclusions_future_wotk} provide conclusions of this work and highlights other research directions for applying DL and DRL to quantum computing.

\section{Background}
\label{sec:background}

\subsection{Quantum Circuit Mapping for Multi-Core Architectures}
Before executing quantum circuits, they must be adapted to the hardware's specific limitations. This process is known as compilation, and mapping constitutes a critical phase in this process. Mapping quantum circuits is significantly challenging when dealing with multi-core quantum computers. This study is based on the assumption of an EPR-distributed architectural model, focusing on the distribution of quantum states across cores. Throughout the circuit's execution, quantum states are dynamically moved from one core to another, ensuring that whenever a two-qubit gate is to be executed, the qubits involved are located within the same core.

To this end, we define a timeslice as a grouping of quantum gates from the circuit that can be executed in parallel. In each timeslice, the interacting qubits are situated in the same core, guaranteeing the feasibility of all two-qubit gates. The movements of qubits between cores are performed by using an EPR-based communication primitive called quantum teleportation, which will be carried out between timeslices. This teleportation protocol is quantum-coherence, thereby enabling the transmission of arbitrary quantum states \cite{remote}. We refer to these inter-core movements as \textbf{non-local communications}.

\begin{figure}[htbp]
\vspace{-0.3cm}
\centerline{\includegraphics[width=\columnwidth]{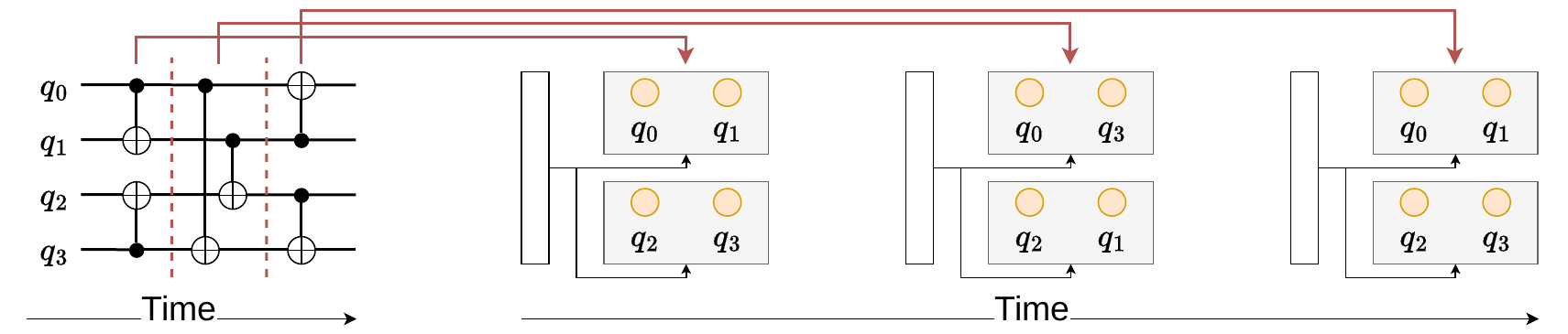}}
\vspace{-0.2cm}
\caption{Quantum circuit (timeslices in red) mapping into a modular architecture. The lower row shows how the mapping can be approached as a graph-partitioning problem.}
\label{fig:multi-core mapping}
\end{figure}

Non-local communications are considerably more costly than intra-core operations. Circuit mapping needs to ensure that, for every two-qubit gate operation, the involved qubits are placed within the same core, illustrated in Figure \ref{fig:multi-core mapping}, and has been the subject of prior research efforts in \cite{chong, qubo, hqa}. While the chosen architectural model supports remote gate execution \cite{remote}, this work exclusively considers qubit movement across cores, as in \cite{chong, qubo, hqa}.

\vspace{-0.2cm}

In \cite{chong}, it is proposed to solve the circuit mapping problem as a graph partitioning approach. For each timeslice, a graph from the interactions of qubits is created, weighting the edges based on the immediacy of the interaction. Then, a graph partitioning algorithm called relaxed Overall Extreme Exchange \cite{partitioning} splits the graph into $k$ clusters (cores), distributing, for each timeslice, all qubits into different cores. This process is repeated for each timeslice, obtaining a path of assignments of qubits to cores.

In \cite{qubo}, a graph-partitioning approach is also proposed and solved using unconstrained quadratic optimization. On the other hand, in \cite{hqa}, a different approach is proposed, obtaining the mapping as an assignment of operations into cores.

This work is analogous to the one proposed in \cite{chong}, approaching the mapping as a graph partitioning problem. Therefore, the FGP-rOEE algorithm proposed in \cite{chong} will be the baseline for this work.
% We aim to obtain similar results than FGP-rOEE, while using a Deep Reinforcement Learning-based graph partitioning algorithm.

\subsection{Deep Reinforcement Learning}
Deep Reinforcement Learning (DRL) is an area of machine learning that combines deep learning techniques with reinforcement learning algorithms \cite{drl}, enabling agents to learn and make decisions in complex environments. Specifically, neural networks approximate the Policy Network and the Value Function. This field has gained significant attention in recent years due to its capacity to address various challenging problems, notably in game-playing and robotics \cite{drl2}.

Since there is no dataset available for optimally mapping quantum circuits onto modular architectures (due to the graph partitioning problem being NP-hard), and current solutions rely on heuristics \cite{chong, qubo, hqa}, traditional deep learning techniques are not a viable option. On the other hand, DRL comes as an alternative solution because it does not require an optimal dataset and has the potential to outperform heuristic-based algorithms

\subsubsection{Proximal Policy Optimization}

Proximal Policy Optimization (or PPO) \cite{ppo} is a widely used algorithm in the field of DRL. It is designed to provide stable and reliable learning by using a trust region approach that restricts the size of policy updates. This ensures that the policy changes are not too large, preventing instability during the learning process. PPO is also known for its sample efficiency, as it efficiently reuses and reweights collected samples to make the most of the available data.

\subsubsection{MaskablePPO}

MaskablePPO \cite{maskppo}, a novel variant of the Proximal Policy Optimization (PPO) algorithm, introduces a masking mechanism to enhance its capabilities. This variant extends the core PPO algorithm by incorporating a mask that selectively determines which parts of the policy should undergo updates during training. This fine-grained control provided by MaskablePPO enhances adaptability and control during the training process and achieves faster convergence.

The fundamental mechanism behind MaskablePPO is straightforward yet effective. By selectively suppressing certain actions based on their validity in specific states, the algorithm can learn policies that closely align with desired behaviours or constraints. Such a level of granularity provides valuable flexibility when dealing with complex environments.

\section{Deep Reinforcement Learning Exploration} \label{sec:DRL_Exploration}

The ultimate goal of the DRL agent is, for a given quantum circuit segmented into timeslices, to find valid assignments (of qubits into cores) while minimizing the average non-local communications during the circuit execution.

The observation of the agent plays a crucial role in the agent's behaviour, to the point that the decision of what action to perform is based purely on the observation array, which, in DRL, is the input of both value function and policy networks.

A key feature of the FGP-rOEE algorithm is the incorporation of lookahead weights. These weights encode the interaction of qubits in future timeslices and guide the swapping decisions. Consequently, the agent, apart from observing the interacting qubits for the current timeslice, can also consider future interactions.

This work's observation array comprises several components, including the lookahead weights and the current and last assignment (qubits to cores). The last assignment refers to the mapping of qubits to cores obtained in the previous timeslice, which is used as the initial assignment for the current timeslice. Finally, a binary value is also included, with a value of 1 when all interacting qubits are placed in the same core, thus obtaining a valid assignment and 0 otherwise.

The action space, another key element in defining a DRL model, defines how the agent interacts with the environment, and choosing the right actions leads to successful agent performance. Following the approach used in the state-of-the-art algorithm FGP-rOEE, the actions used by the agent are performing swaps between two qubits, i.e. exchanging the location (core) of two qubits.

The agent's primary objective is to find valid assignments for each timeslice sequentially. The environment setup requires the agent to find a valid assignment within a limited number of actions before progressing to the subsequent timeslice. Once there, the agent begins to find a new assignment, using the previous assignment as a base.
% The initial mapping of qubits to cores is done randomly, satisfying the constraint of maintaining an equal number of qubits in each machine.

The reward function has been defined as follows: for each valid assignment found, the agent receives a reward based on the number of non-local communications required in between the previous and the current timeslice. Once a valid assignment has been found for all timeslices, a final reward is given based on the average number of non-local communications.

All the approaches we will discuss use the same observation, action space, and reward function. This aims to examine how various models perform under different mask conditions. To assess whether the DRL models are experiencing significant learning, we will compare each of them to an identical model, having the same settings but prior to any training, which we refer to as a random execution.

\subsection{Approach 1: PPO} \label{subsct:app1}
The fundamental idea behind this approach is to grant the agent absolute freedom, allowing it to swap between all qubits indiscriminately. This design encourages the agent to explore a wider set of possibilities, potentially resulting in the discovery of efficient strategies for qubit allocation and a reduction in the average number of non-local communications.

When comparing the performance of the trained agent with a random execution, it was observed that the agent successfully achieved an average reduction of 36\% in non-local communications.

\subsection{Approach 2: Soft Mask} \label{subsct:app2}
This approach is centred on analysing an agent's behaviour when certain actions, which we term \textit{useless actions}, are disabled. These actions are disabled by implementing a masking mechanism through the MaskablePPO algorithm, which removes the masked actions from the action space. Actions falling into this category include swapping identical qubits, swapping two qubits that are allocated in the same machine, or proceeding to the following timeslice without achieving a valid assignment for the current one. We anticipate a quicker convergence and more efficient behaviour by removing these actions from the agent's action space.

The implementation of this approach resulted in a 30\% reduction in non-local communications compared to a random execution where the same mask was applied.

\subsection{Approach 3: Hard Mask} \label{subsct:app3}

The central concept of this approach is that by imposing a more restrictive mask and guiding the agent towards a good solution, the model will be able to find better assignments for each timeslice. The mask assigned to the agent was designed with the main objective of reducing the average number of non-local communications, implementing a \textit{direct-swap} heuristic on top of the mask presented in the last approach. The \textit{direct-swap} heuristic only considers the movements of misplaced qubits to the core, which it needs to interact with.

This mask imposes a large limitation on the actions available in every state, guiding the agent towards a good solution but restricting its ability to learn deeper underlying patterns. This causes the trained model and the random execution (the same mask was applied) to have similar results on the number of non-local communications. The lack of improvement in the trained model indicates that the mask reduces the learning capabilities of the agent.

\subsection{DRL Models Discussion}

In this section, we provide a comparative analysis of the previously discussed approaches. The analysis focuses on the effectiveness in achieving the goal, along with evaluating key metrics such as the evolution of the reward function and the length (number of actions taken by the agent) of each episode.

Though the agent's primary goal is to reduce the number of non-local communications, it is also important to monitor the reward function's progression, which indicates the agent's learning progress. Additionally, the duration of each episode is a factor to consider, given its direct correlation with inference time.

\begin{figure*}[htbp]
\vspace{-0.3cm}
\centerline{\includegraphics[width=\textwidth]{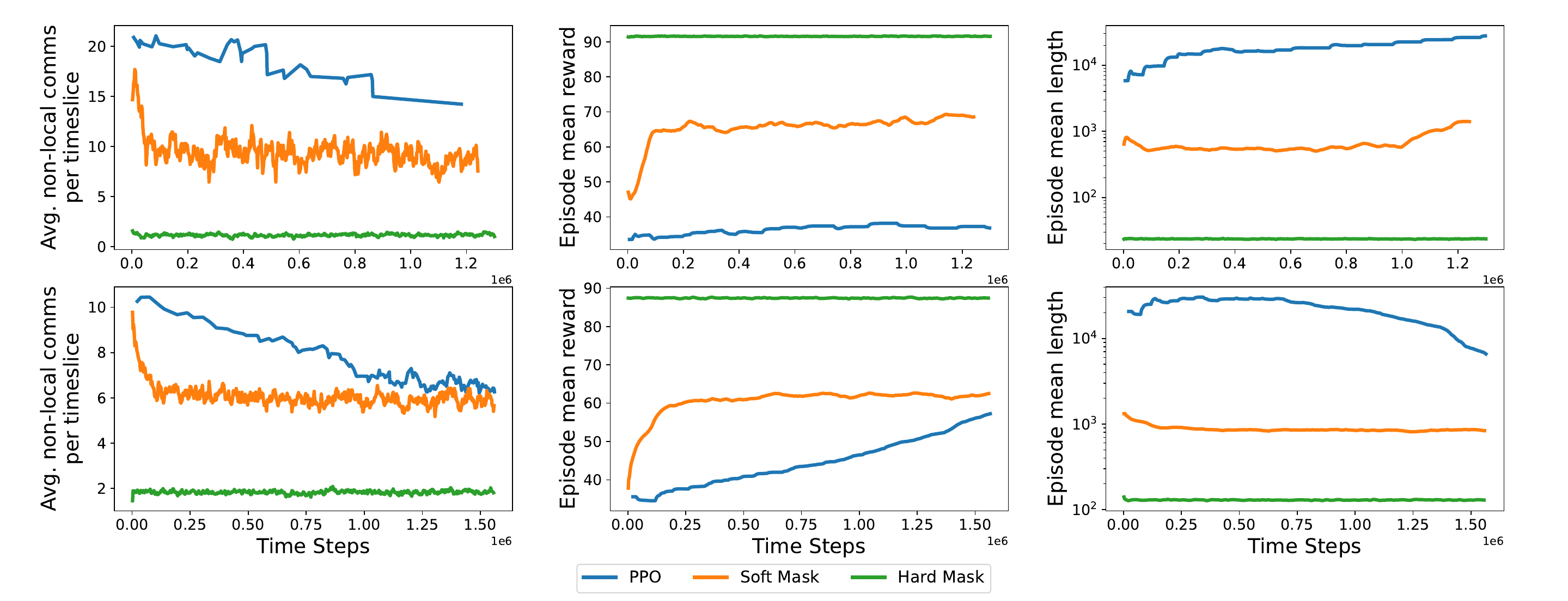}}
\vspace{-0.2cm}
\caption{Model performance within a 16 (upper row) and 32-qubit (lower row) modular quantum computer with four cores. Columns correspond to the evaluation metrics, from left to right, non-local communications, mean reward, and mean length.}
\label{fig:model-comp}
\end{figure*}

The left and right columns of Figure \ref{fig:model-comp} illustrate how limiting the action space results in better performance, both in the number of non-local communications and the length of each episode.

Upon closely comparing the PPO and the Soft Mask approaches, it can be observed that the optimal performance of both yields very similar results since the mask applied in the second approach only limits actions that do not impact the observation (the mask does not guide the agent in any direction). Therefore, we can see that both approaches can achieve roughly the same performance, with the second approach converging much faster (middle column in Figure \ref{fig:model-comp}), meaning that the PPO model learns to mimic the mask used in the Soft Mask approach. However, in terms of episode mean length, it can be seen how the Soft Mask approach obtains a much lower value than the PPO model, resulting in faster inference time.

% deduced by the similar results and because, in the top left plot in Figure \ref{fig:multi-core mapping}, the length of the episodes decreases over time, meaning that the agent learns to be more efficient.

% It’s also important to note that even though they both achieve similar results, the number of actions taken in the first approach is almost an order of magnitude bigger (left column \ref{fig:multi-core mapping})

Regarding the Hard Mask approach, the right column in Figure \ref{fig:model-comp} highlights its substantial superiority in reducing the number of non-local communications, setting it apart from the other approaches. However, a closer look at the middle column in Figure \ref{fig:model-comp} reveals that the reward curve suggests a lack of evolution in the learning process, evidenced by the consistent reward obtained across all timesteps. Our insights are that the restrictive mask employed, which considerably constrains the feasible actions in each state, steers the agent towards a better solution than the two other approaches while limiting the agent’s capacity to discern more complex strategies that could potentially enhance performance.

\section{Performance Assessment} \label{sec:Performance_Assessment}

In this section, we delve into a comprehensive evaluation of the various proposed DRL models. We only consider those that use masks, as they yield the best results. This section's primary objective is to compare the efficacy of our models with the state-of-the-art FGP-rOEE algorithm.

%To ensure a robust and unbiased comparison, we employ a diverse range of circuits, varying not only in size but also in type. This variety allows us to assess the performance across a broad spectrum of scenarios, thereby providing a more holistic view of the strengths and potential areas for improvement.

% \begin{figure}[!htbp]
% \vspace{-0.3cm}
% \centerline{\includegraphics[width=0.85\columnwidth]{figures/performance_assessment/result_ratio_chong.pdf}}
% \vspace{-0.2cm}
% \caption{Ratio between the state-of-the-art FGP-rOEE algorithm and the different maskable approaches}
% \label{fig:ratio_chong}
% \end{figure}
% \vspace{-0.2cm}

\begin{figure}[!htbp]
\vspace{-0.3cm}
\centerline{\includegraphics[width=\columnwidth]{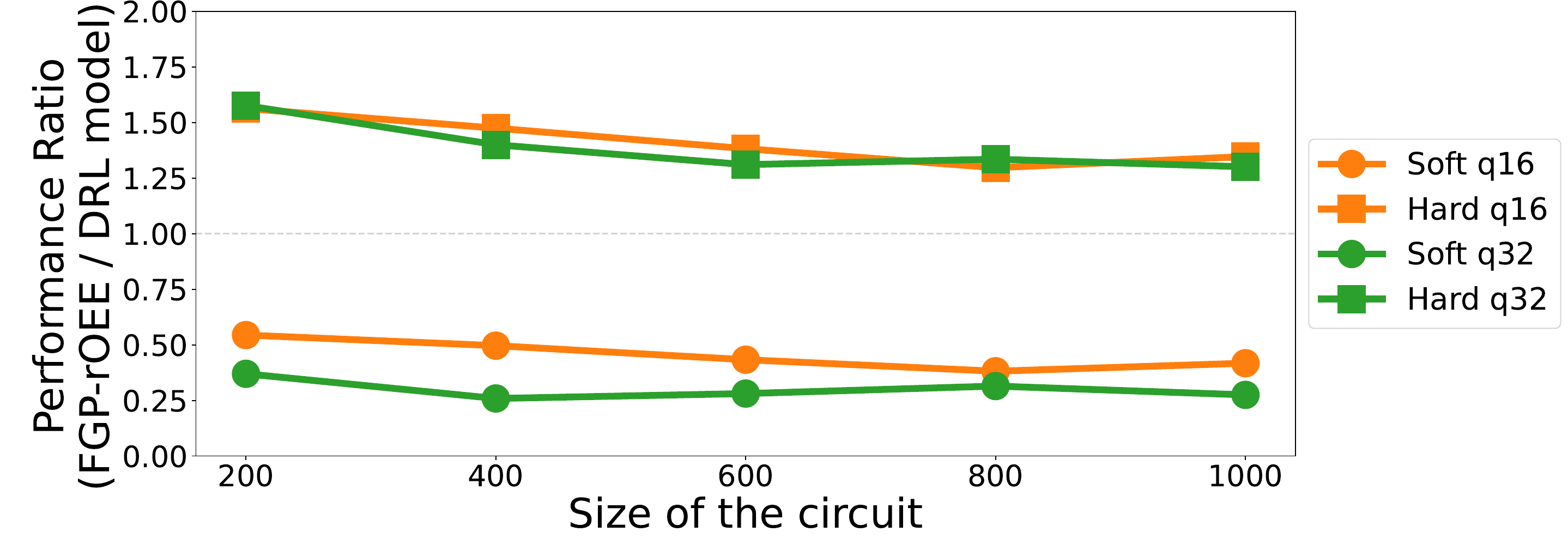}}
\vspace{-0.2cm}
\caption{Ratio between the state-of-the-art FGP-rOEE algorithm and the different maskable approaches}
\label{fig:ratio_chong}
\end{figure}

Figure \ref{fig:ratio_chong} depicts the ratio of non-local communications between the proposed models and the state-of-the-art. Points above one indicate that the model outperforms the state-of-the-art algorithm, while points below one indicate the opposite. These results demonstrate that the Hard Mask consistently outperforms FGP-rOEE. On the other hand, for the Soft Mask, the models are unable to match the state of the art and fall behind, obtaining a higher number of non-local communications.

In our comparative study, we benchmarked the same models with other types of quantum circuits, such as Cuccaro \cite{cuccaro} and QAOA \cite{qaoa}. The results were consistent across both circuit types. In both cases, with 32 qubit circuits, the Hard Mask approach demonstrated a slight edge over FGP-rOEE, yielding 1.20$\times$ better results with the first circuit and 1.05$\times$ with the latter. On the other hand, for the Soft Mask, the results did not fare as well. The performance dropped significantly, yielding ratios of 0.14$\times$ and 0.17$\times$, respectively.
% In both cases, the results are inferior to those obtained from random circuits. However, it is worth noting that both models were exclusively trained with random circuits. Therefore, it's plausible that these models could potentially perform better if they were trained specifically with these types of circuits or a combination of them. \textcolor{red}{Pau: Be careful, what does it mean to be trianed with specifically these types of circuits.}

The key takeaway from this section is that we got evidence that the models generalize well and consistently outperform the state-of-the-art algorithm in a diverse range of circuits.

\section{Conclusions and Future Work} \label{sec:Conclusions_future_wotk}

This paper presents a novel approach to partitioning algorithms for quantum applications using DRL. The Hard Mask model outperforms existing algorithms by minimizing non-local communications and shows potential for use in various real-world scenarios. Despite not being as successful as the Hard Mask, proposed models like PPO or Soft Mask show signs that DRL can solve this problem as the agents can learn to reduce non-local communications.

Future work should focus on achieving more pronounced learning curves to ensure the agent is learning effectively. This could involve adjusting the reward system, refining the observation, or expanding the range of the action space. Experimenting with different neural network architectures and better hyperparameter tuning could also help optimize the DRL model. Additionally, it would be beneficial to train the DRL model on larger circuits using more powerful computing resources to test its scalability and performance under demanding conditions.

\newpage
\balance

\end{document}